\documentclass[12pt]{elsarticle}

\usepackage{amssymb}
\usepackage{color}
\usepackage{threeparttable}
\usepackage{array}
\usepackage[nodots]{numcompress}
\usepackage{amsmath}
\newcommand{\onlinecite}[1]{\hspace{-1 ex} \nocite{#1}\citenum{#1}}
\newcommand{\PreserveBackslash}[1]{\let\temp=\\#1\let\\=\temp}
\newcolumntype{C}[1]{>{\PreserveBackslash\centering}p{#1}}
\newcolumntype{R}[1]{>{\PreserveBackslash\raggedleft}p{#1}}
\newcolumntype{L}[1]{>{\PreserveBackslash\raggedright}p{#1}}
\usepackage{caption}
    \makeatletter
    \def\ps@pprintTitle{%
      \let\@oddhead\@empty
      \let\@evenhead\@empty
      \def\@oddfoot{\reset@font\hfil\thepage\hfil}
      \let\@evenfoot\@oddfoot
    }
    \makeatother

\begin{document}
\begin{frontmatter}



\title{Enhancing Composition Window of Bicontinuous Structures by Designed Polydispersity Distribution of ABA Triblock Copolymers}


\author{\mbox{Yue Li}}
\author{\mbox{Hu-Jun Qian}\footnote{Corresponding author. E-mail: hjqian@jlu.edu.cn}}
\author{\mbox{Zhong-Yuan Lu}\footnote{Corresponding author. E-mail: luzhy@jlu.edu.cn}}

\address{State Key Laboratory of Theoretical and Computational
Chemistry, Institute of Theoretical Chemistry,Jilin University,
Changchun 130023, China}

\author{\mbox{An-Chang Shi}\footnote{Corresponding author. E-mail: shi@mcmaster.ca}}

\address{Department of Physics and Astronomy, McMaster University, Hamilton, Ontario, Canada L8S 4MI}

\begin{abstract}
The phase behavior of polydisperse ABA triblock copolymers is studied using dissipative particle dynamics simulations, focusing on the emergence and property of bicontinuous structures. Bicontinuous structures are characterized by two separate, intermeshed nanoscopic domains extending throughout the material. The connectivity of polymeric bicontinuous structures makes them highly desirable for many applications. For conventional monodisperse diblock and triblock copolymers, regular bicontinuous structures (\textit{i.e.}, gyroid and Fddd) can be formed over a narrow composition window of $\sim$ 3$\%$. We demonstrate that the composition window  for the formation of bicontinuous structures can be regulated by designed polydispersity distributions of ABA triblock copolymers. In particular, introducing polydispersity in both A and B blocks can lead to a significant enhancement of the composition window of bicontinuous structures with {\em both} continuous A and B domains. The mechanism of the bicontinuous structure enhancement is elucidated from the distribution of the long and short blocks. Furthermore, it is shown that the polymeric bicontinuous structures from polydisperse ABA triblock copolymers possess good continuity throughout the sample, making them ideal candidates for advanced applications.
\end{abstract}


\end{frontmatter}

% \linenumbers

\section{INTRODUCTION}
\label{}
\par
Block copolymers have received considerable attention in the past decades due to their inherent ability  to spontaneously form ordered morphologies at length-scales spanning from less than a few nanometers to a few hundreds nanometers~\cite{bates_block_1999}. By precisely tuning block type and architecture, extraordinary control over the self-assembled ordered structures can be achieved~\cite{matyjaszewski_architecturally_2011,bates_multiblock_2012}. The advances in morphological control enable diverse and expanding applications of block copolymers in areas such as drug delivery~\cite{kataoka_block_2001}, solar cells~\cite{ho_poly3-alkylthiophene_2011}, and nanolithography~\cite{segalman_directing_2008,ruiz_density_2008,chai_scanning_2010}.
\par
To date, linear diblock and triblock copolymers are the two simplest model systems for the study of block copolymer self-assembly both theoretically and experimentally. It is well established that the phase behavior of monodisperse diblock/triblock copolymers is mainly determined by the incompatibility between the distinct blocks (characterized by \emph{$\chi$N}, where \emph{$\chi$} is the Flory-Huggins parameter and \emph{N} the chain length) and molecule composition ($f$). Furthermore, it has been demonstrated conclusively that polydispersity in the block length can have a great impact on the phase behavior of block copolymers~\cite{register_materials_2012,widin_unexpected_2012,lynd_influence_2005,Lynd2008875}.
For instance, in the case of linear AB diblock copolymers, it has been found that polydispersity can (i) expand the domain spacing of ordered structures; (ii) shift the order-order transition boundaries (OOTs) towards larger volume fraction of polydisperse component; and (iii) alter the order-disorder transition point, $(\emph{$\chi$N})_{\text{ODT}}$, i.e., $(\emph{$\chi$N})_{\text{ODT}}$ decreases when the polydisperse blocks form the minority domain, whereas it increases when the polydisperse blocks form the majority domain~\cite{lynd_effects_2007}.
For the case of linear ABA triblock copolymers, Mahanthappa and coworkers reported unexpected consequences of block length polydispersity on the self-assembly process~\cite{widin_unexpected_2012}. Specifically these authors observed that poly(styrene-$b$-1,4-butadiene-$b$-styrene) (SBS) triblock copolymers with middle B-blocks having a broad length distribution can also self-assemble into conventional ordered structures as the monodisperse analogs. At the same time, molecular weight polydispersity in the center B-blocks causes large shifts of order-order transition boundaries between the composition windows. These observations have been confirmed in a recent self-consistent field theory (SCFT) calculation by Matsen~\cite{Matsen_EPJ2013}. Furthermore, the experiments by Mahanthappa {\it et al.} reveal that a highly desirable intermeshed bicontinuous structure could be stabilized over a relatively broad composition window of the middle blocks ($f_{B}\sim 10\%$)~\cite{widin_unexpected_2012}. As pointed out by Register~\cite{register_materials_2012}, such a bicontinuous structure is highly preferable for many applications such as photovoltaic films in solar cells~\cite{crossland_bicontinuous_2009} and separator membranes in batteries~\cite{singh_effect_2007}, where a nanoscopic bicontinuous structure could facilitate electron excitation and ion diffusion.
\par
From a practical point of view, the following experimental observations~\cite{widin_unexpected_2012, widin_polydispersity-induced_2010} are particularly important: (i) Bicontinuous gyroid structures formed by monodisperse diblock/triblock copolymers only exist in a very narrow composition window of $\sim$ 3$\%$, whereas polydisperse triblock copolymer can form bicontinuous network structures in a much larger composition window of $\sim$ 10$\%$; (ii) The polydisperse middle B-block can be synthesized via much cheaper methods. However, as pointed out by Register~\cite{register_materials_2012}, some fundamental questions on polydispersity induced bicontinuous structures still remain: (i) Is asymmetric polydispersity necessarily required? That is, whether the narrowly distributed lengths of the A-blocks are required? Or practically, whether the end A-blocks can also be synthesized via cheaper methods? (ii) Are both domains fully continuous across the entire composition range for which the irregular bicontinuous structure forms? (iii) In a polydisperse system, some very short B-blocks are readily dissolved in the A-domain. How does the mixing of A and B blocks affect the physical properties of the material (such as glass transition temperature and mechanical integrity of glassy domains)?
\par
Motivated by these recent progresses and questions, we carried out extensive dissipative particle dynamics (DPD) simulations~\cite{groot_dissipative_1997} to investigate the effect of polydispersity on the phase behavior of ABA triblock copolymers. In our previous study on diblock copolymers~\cite{liyue_diblock}, we have successfully adopted the DPD method to reproduce quantitatively the polydispersity induced increase in lamellar domain spacing and clarified the underlying molecular mechanism. Our main aim in the current contribution is to address not all but at least some of the aforementioned questions, mainly the necessity of the narrow distribution in A-blocks, and the continuity of the irregular bicontinuous structure. To this end, we focus on three representative ABA triblock copolymer systems: (1) all the blocks are monodisperse; (2) only the middle B-block is polydisperse; and (3) both A and middle B blocks are polydisperse. In particular, the systems with large polydispersity in the middle B-block show very good agreement with experiments~\cite{widin_unexpected_2012} on the positions of composition windows of ordered domains. In systems with polydispersity in both A and B blocks, irregular network structures formed by B-blocks are also obtained, indicating that the narrow distribution of end A-blocks is not a prerequisite of forming bicontinuous morphologies. More importantly, the end A-blocks with polydispersity are found to have the same probability as middle B-blocks to form irregular network structures when the volume fraction of A component is small. Therefore the composition window for fabricating highly desirable irregular bicontinuous structures can be enlarged to $\sim$ 20$\%$ by introducing polydispersity into both end A and middle B blocks. The stabilization of irregular bicontinuous structures can be attributed to the reduction of molecular packing frustration due to the selective distribution of long blocks in the interior of the bicontinuous network structure nodes. These results shed new light on understanding the chain length polydispersity effect on block copolymer microphase separation. It also suggests that instead of using expensive living anionic, cationic, or metal-catalyzed methods, much economical radical polymerization method may be used to synthesize ABA triblock copolymers for the fabrication of bicontinuous structures.
%%%%%%%%%%%%%%%%%%%%%%%%%%%%%%%%%%%%%%
\section{COMPUTATIONAL DETAILS}
\subsection{Dissipative particle dynamics (DPD) method}
The DPD method is a well established particle-based method and it has been successfully applied to investigate block copolymer phase separations~\cite{groot_dissipative_1997,groot_dynamic_1998,qian_dissipative_2005}. In a DPD model, a group of adjacent segments is represented by a coarse-grained bead. Polymers are described by a bead-spring model with adjacent coarse-grained beads connected by harmonic spring forces $\textbf{F}^{S}$ = $-k_s\textbf{r}_{ij}$, in which the force constant $k_s$ is set to 4.0 in reduced units according to Ref.~\onlinecite{groot_dynamic_1998}. The force acting on each bead is pairwise additive, and is composed of a conservative force $F^{C}$, a dissipative force $F^{D}$, and a random force $F^{R}$. All the forces act along the connecting line of interacting bead centers, with
\begin{eqnarray}
\mathbf{F}_{i}&=&\sum_{j\neq i}(\mathbf{F}^{C}_{ij} + \mathbf{F}^{D}_{ij} + \mathbf{F}^{R}_{ij} + \mathbf{F}^{S}_{ij} )\nonumber,\\
\mathbf{F}_{ij}^C&=&\alpha_{ij}\omega^C\left(r_{ij}\right)\mathbf{e}_{ij}\nonumber,\\
\mathbf{F}_{ij}^D&=&-\gamma\omega^D\left(r_{ij}\right)\left(\mathbf{v}_{ij}\cdot\mathbf{e}_{ij}\right)\mathbf{e}_{ij}\nonumber,\\
\mathbf{F}_{ij}^R&=&\sigma\omega^R\left(r_{ij}\right)\xi_{ij}\Delta{t}^{-1/2}\mathbf{e}_{ij},
\end{eqnarray}
where $\mathbf{r}_{ij} = \mathbf{r}_{i} - \mathbf{r}_{j}$; $r_{ij} = |\mathbf{r}_{ij}|$ and $\mathbf{e}_{ij} = \mathbf{r}_{ij}/r_{ij}$. $\textbf{v}_{ij}$ is the relative velocity between interacting beads $i$ and $j$. $\alpha_{ij}$ is the maximum repulsion parameter between bead $i$ and bead $j$, $\gamma$ and $\sigma$ are the amplitudes of the friction and the noise strength, respectively. $\omega^C(r_{ij})$, $\omega^D(r_{ij})$ and $\omega^R(r_{ij})$ are weight functions. The dissipative and random forces are coupled together to form an effective thermostat according to fluctuation-dissipation theorem~\cite{espanol_statistical_1995} with the relations of $\sigma^{2}=2\gamma k_{B}T$ and $\omega^D(r_{ij})=[\omega^R(r_{ij})]^{2}$. A simple form of $\omega^C(r_{ij})=\omega^R(r_{ij})=(1-r_{ij}/r_c)$ (when $r_{ij}\leq r_{c}$ and otherwise equals to 0) is chosen in our simulations. $\xi_{ij}$ is a random variable with zero mean and unit variance. For easy numerical handling, we have chosen the cutoff radius $r_c$, the bead mass $m$, and $k_{B}T$ as the units in the simulations, so the unit of time $\tau$ is $\tau = r_{c} \sqrt{m/k_{B}T} =1 $.
\par
The bead density is set at $\rho$ = 3 in our simulations. The time evolution of the interacting beads is governed by Newton's equations of motion. DPD-VV algorithm~\cite{groot_dissipative_1997} is used to propagate the system with a time step of $\Delta{t}$ = 0.02. Each simulation is 3$\times$$10^6$ $\sim$ 6$\times$$10^6$ steps long and five parallel simulations have been done for each thermodynamic condition in the phase diagram.
\par
\subsection{Calculation of Flory-Huggins $\chi$ parameter}
The conservative interaction strength $\alpha_{ij}$ in DPD can be mapped to Flory-Huggins \emph{$\chi$} parameter using the recipe proposed by~\cite{groot_dissipative_1997}:
\begin{eqnarray}
\label{eq:alphachi}
\alpha_{ij}=\alpha_{ii}+3.27\chi \quad (\rho=3),
\end{eqnarray}
where $\alpha_{ii}$ is the interaction strength between the beads with the same type and it takes the value of 25.0 for $\rho$ = 3 to represent typical incompressible fluid.
\par
According to the theoretical calculations including fluctuation effect~\cite{mayes_concentration_1991}, order-disorder transition at composition fraction $f$ = 0.5 for block copolymers with finite chain length will shift from the mean field prediction of $(\emph{$\chi$N})_{c}$ as:
\begin{eqnarray}
\chi N=(\chi N)_c+\lambda \bar{N}^{-1/3},
\end{eqnarray}
where $(\emph{$\chi$N})_{c}$ = 17.9 is the order-disorder transition (ODT) value for monodisperse ABA triblock copolymers predicted by mean field theory and $\lambda$ is a positive constant to be determined for triblock copolymers, $\bar{N}$ = $6^{3}(R_{g}^{3}\rho_{p})^{2}$ is the invariant chain length, $R_{g}$ is the radius gyration and $\rho_{p}$ is the polymer concentration.  Similar to diblock copolymers~\cite{groot_dynamic_1998}, effective $\chi$-parameter in simulations for triblock copolymer can be rescaled by a factor:
\begin{eqnarray}
\label{eq:chieffective}
(\chi N)_{eff}=\frac{17.9}{17.9+\lambda \bar{N}^{-1/3}}\chi N=\frac{1}{1+C\bar{N}^{-1/3}}\chi N,
\end{eqnarray}
where the constant $C$ = $\lambda/17.9$. Considering the fact that the polymers with finite chain lengths in simulations usually deviate from Gaussian, i.e., the radius of gyration of the chain $R_g$ obeys a scaling of $R_g$ $\sim$ $ N^\nu$, Eq.(\ref{eq:chieffective}) thus can be modified as~\cite{liyue_diblock,chen_effects_2005,Qian_cyclic_2005}:
\begin{eqnarray}
\label{eq:chieff}
(\chi N)_{eff}=\frac{1}{1+C\bar{N}^{2/3-2\nu}}\chi N,
\end{eqnarray}
where $\nu$ is the swelling exponent which can be determined by measuring the radius of gyration of polymers in the simulations~\cite{liyue_diblock,chen_effects_2005,Qian_cyclic_2005}, $\langle R^{2}_{g}\rangle$ = 1/6$N^{2\nu}l^{2}$. To estimate $C$ in Eq.(\ref{eq:chieff}), we simulate a calibrating monodisperse triblock copolymer system A$_{3}$B$_6$A$_3$ with a stepwise change in the interaction parameter $\alpha$ from 37 to 40 with $\Delta\alpha$ = +0.5. The results show that the system is in the disordered melting state below $\alpha$ = 39.0 and the ordered lamellar structure appears above $\alpha$ = 39.5. Therefore we take $\alpha$ = 39.5 as the interaction strength at the order-disorder transition point. According to Eq.(\ref{eq:alphachi}), a value of 53.2 for $(\emph{$\chi$N})_{\text{ODT}}$ can be readily obtained. A further mapping using Eq.(\ref{eq:chieff}) to the mean field value of $(\emph{$\chi$N})_{\text{ODT}}^{eff}$ = 17.9 immediately gives the $C$ value of 5.36. Therefore, in order to directly compare our simulation results with SCFT and experimental results, all the segregation strengths are mapped onto $(\emph{$\chi$N})_{eff}$ using Eq.(\ref{eq:chieff}) with $C$ = 6.485 in the following simulations.
\par
\subsection{Polydisperse models}
In this study, monodisperse or polydisperse triblock copolymer $\text{A}_{x1}\text{B}_{N-x1-x2}\text{A}_{x2}$ melts ($f_{\text{A}}$ = $(x1+x2)$/$N$ and $f_{\text{B}}$ = 1-$f_{\text{A}}$) with average chain length of $N$ = 8 $\sim$ 18 are simulated in a simulation box with a volume of 40$\times$40$\times$40. The molecular characteristics for monodisperse and polydisperse ABA triblock copolymer are given in Tables~\ref{tb1:sample1},~\ref{tb2:sample2},~\ref{tb3:sample3}, and~\ref{tb4:sample4}, respectively. In the monodisperse systems (Table~\ref{tb1:sample1}) or in the systems with polydispersity only presented in middle B-blocks (Table~\ref{tb3:sample3}), all block copolymer chains are symmetric, i.e., $x1$ = $x2$. In systems with both A and B blocks polydisperse (Table~\ref{tb2:sample2} and Table~\ref{tb4:sample4}), the end A-blocks are connected to middle B-block in a random way, mimicking the situation in a radical polymerization process, therefore the symmetry of the polymer chains is broken.
\par
In order to represent the wide molecular weight distribution in radical polymerization and narrow distribution in anionic polymerization processes, Schulz-Zimm and Poisson distributions are utilized, respectively, to construct our polydisperse triblock copolymer systems. The Schulz-Zimm distribution obeys the form of~\cite{zimm_apparatus_1948,lynd_theory_2008}:
\begin{eqnarray}
p(N)=\frac{u^{u}\delta^{u-1}\exp(-u\delta)}{N_n\Gamma(u)},
\end{eqnarray}
where $\delta$=$N$/$N_n$, $N_n$ is the average chain length. The polydispersity index (PDI) for this distribution is PDI=$(u+1)/u$.
\par
The Poisson distribution is specified by:
\begin{eqnarray}
p(N)=\frac{\exp(1-N_n)(N_n-1)^{N-1}}{(N-1)!}.
\end{eqnarray}
The PDI for Poisson distribution is PDI = 1 + $(N_n-1)/N_{n}^2$.
\par
%In order to facilitate the comparison between existing experimental results, polydispersity is considered in block B, namely the polydisperse B block has a segmental length of $f_{\text{B}}$$N$$\sigma$, where $\sigma$ is a variable with the well-known Schulz-Zimm distribution~\cite{zimm_apparatus_1948,lynd_theory_2008},
%\begin{eqnarray}
%p(\sigma)=\frac{k^{k}\sigma^{k-1}\exp(-k\sigma)}{\Gamma(k)}
%\end{eqnarray}
\subsection{Calculation of the structure factor}
In experiments, scattering function is frequently used to identify the ordered structures via the Bragg reflection ratios $\vec{q}/\vec{q}^*$. For instance, the characteristic ratios for lamellar structure are at $\vec{q}/\vec{q}^*$ = 1, 2, 3, 4, 5, 6, $\cdots$ and 1, $\sqrt{3}$, $\sqrt{4}$, $\sqrt{7}$, $\sqrt{9}$, $\sqrt{11}$,$\cdots$ for hexagonal cylindrical structure~\cite{ryan_ordered_2001}. In our simulations, the minority component domain is always used to calculate the structure factor by:
\begin{eqnarray}
S(\vec{k})&=&\rho_{x}(\vec{k})\rho_{x}(-\vec{k})/N_{x},\nonumber\\
\rho_{x}(\vec{k})&=&\sum^{N_{x}}_{j=1}\exp(i\vec{k}\cdot\vec{r_{j}}^{x}),
\end{eqnarray}
where $N_{x}$ and $\rho_{x}(\vec{k})$ are the bead number and the density of minority component in the reciprocal space, respectively. Wave vector $\vec{q}$ is defined as $\vec{q}$ = (2$\pi/L$)$\vec{k}$, where $L$ is the simulation box size.
\par
\section{RESULTS AND DISCUSSION}
%%%%%%%%%%%%%%%%%%%%%%%%%%%%%%%%%%%%%%%%%%%%%%%%%%%%%%%%%%%%%%%%%%%%%%%%%%%%%
\subsection{Effects of block polydispersity}
\begin{figure}
  \includegraphics[angle=0,width=7cm]{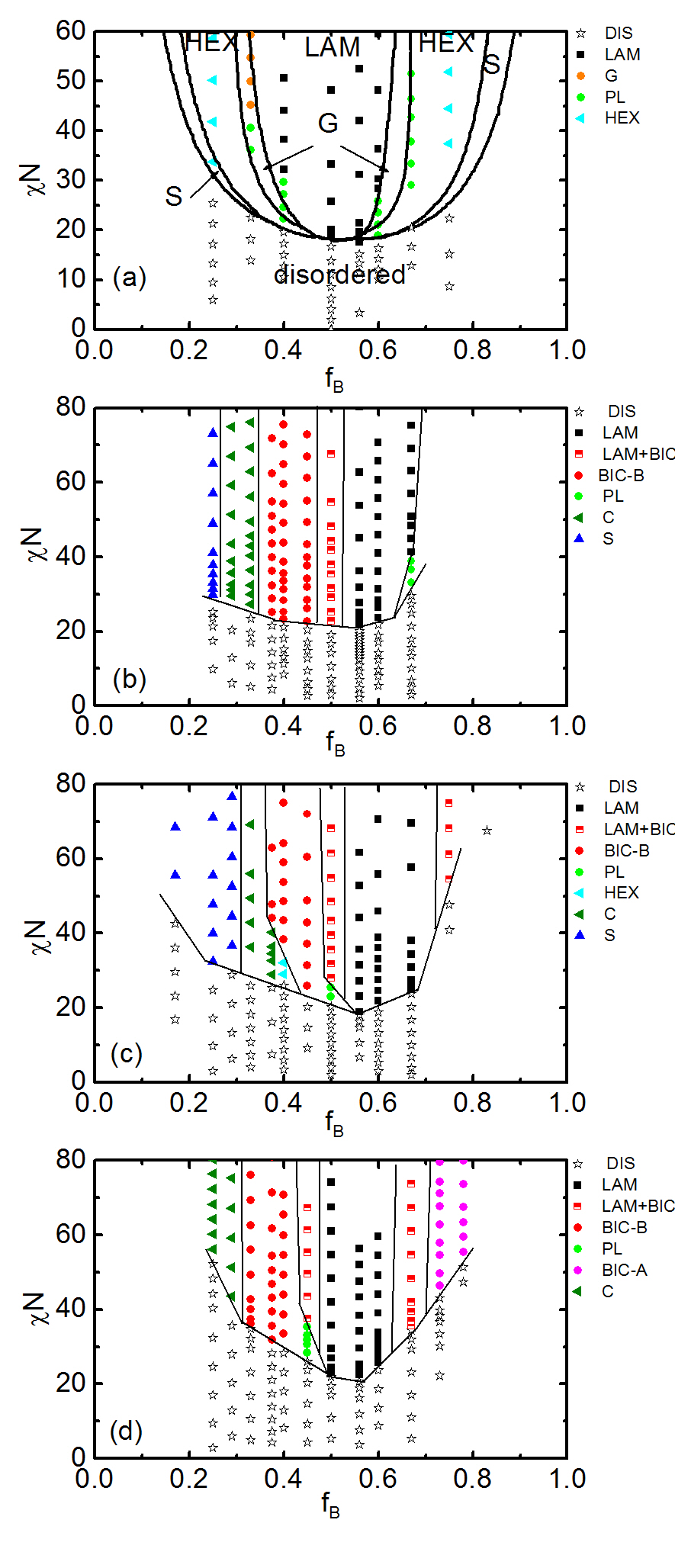}
  \caption{Simulated phase diagrams for ABA triblock copolymers (symbols) (a) in monodisperse system, with the comparison to the SCFT predictions~\cite{matsen_equilibrium_1999} represented by black lines; (b) with PDI$_{\text{B}}$ $\sim$ 2.0 (Schulz-Zimm) and PDI$_{\text{A}}$ $\sim$ 1.25 (Poisson); (c) with intermediate polydispersity in middle B-blocks (PDI$_{\text{B}}$ $\sim$ 1.5, SZ) and monodisperse end A-blocks; (d) with intermediate polydispersity in both A and B blocks (PDI$_{\text{A}}$ $\sim$ PDI$_{\text{B}}$ $\sim$ 1.5, SZ). The morphologies are labeled as DIS: disordered melting phase; S: spherical phase; C: cylinder phase; HEX: hexagonally packed cylinder; G: gyroid; PL: perforated lamellae; LAM: lamellae; BIC-A: bicontinuous network structure formed by A-blocks; BIC-B: bicontinuous network structure formed by B-blocks; LAM+BIC: coexistence of LAM and BIC phases. The microphase boundary lines in (b), (c) and (d) are drawn to guide the eye.}
  \label{fig:FIG1}
\end{figure}

\begin{figure}
  \includegraphics[angle=0,width=14cm]{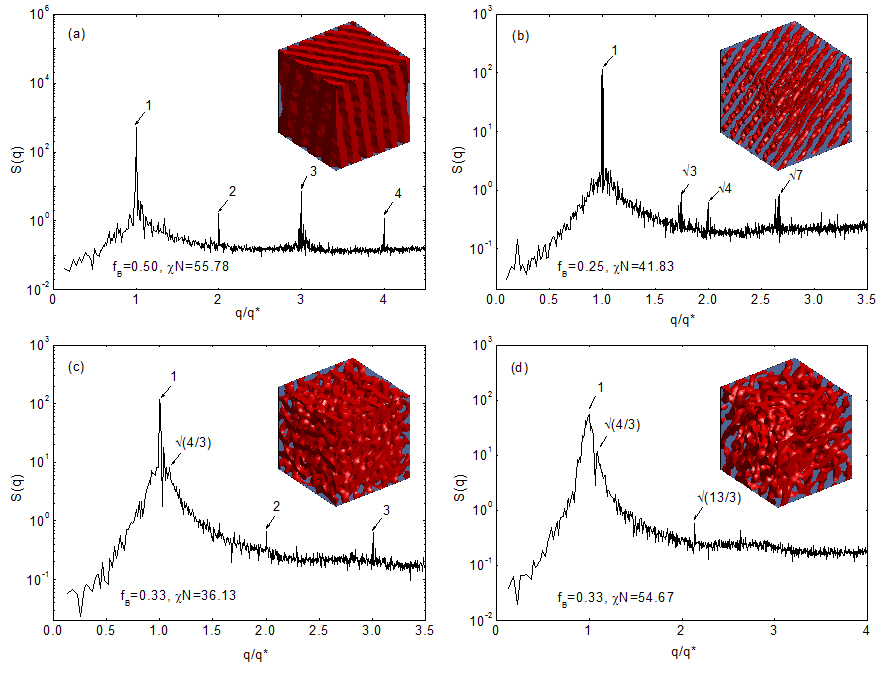}
  \caption{Structure factors and the corresponding mesostructures (insets): (a) lamellae for system mABA-50 at $\chi$N = 55.78; (b) hexagonally packed cylinder for  system mABA-25 at $\chi$N = 41.83; (c) perforated lamellae for system mABA-33 at $\chi$N = 36.13; (d) gyroid phase for system m-ABA-33 at $\chi$N = 54.67.}
  \label{fig:FIG2}
\end{figure}

%%%%%%%%%%%%%%%%%%%%%%%%%%%%%%%%%%%%%%%%%%%%%%%%%%%%%%%%%%%%%%%%%%%%%%%%%%%%%
Figure~\ref{fig:FIG1} illustrates the polydispersity effects on the phase diagram of ABA triblock copolymers. First of all, we compare our simulation results for monodisperse ABA triblock copolymers with SCFT predictions~\cite{matsen_equilibrium_1999} in Figure~\ref{fig:FIG1}(a). In general, we find very good agreement between our simulation results and SCFT predictions. For example, the Lamellar (LAM), gyroid (G), and hexagonally packed cylinder (HEX) structures obtained in our simulations are found to be located in the corresponding composition windows predicted by SCFT. In the SCFT composition window for gyroid structure, we observe some perforated lamellar (PL) structures (green solid circles in Figure~\ref{fig:FIG1}(a)) at relatively low segregation strength, which are typical in DPD simulations~\cite{groot_dynamic_1998}. Isodensity surface plots for lamellar, cylinder, perforated lamellar and gyroid structures and their corresponding structure factors are shown in Figure~\ref{fig:FIG2}, in which Bragg reflection ratios of $\vec{q}/\vec{q}^*$ = 1, 2, 3, 4 are found for lamellar structures (cf. Figure~\ref{fig:FIG2}(a)) and $\vec{q}/\vec{q}^*$ = 1, $\sqrt3$, $\sqrt4$, $\sqrt7$ for cylinder structures (Figure~\ref{fig:FIG2}(b)). These radios are at 1.0, $\sqrt{4/3}$, 2.0, and 3.0 for the perforated lamellar structures (Figure~\ref{fig:FIG2}(c)) and at 1.0, $\sqrt{4/3}$, $\sqrt{13/3}$ for gyroid structures (Figure~\ref{fig:FIG2}(d)). The results of these characteristic ratios are also consistent with available experimental results~\cite{ryan_ordered_2001}.
\par
In order to directly compare simulation results to the experiments of Mahannapa {\it et al.}~\cite{widin_unexpected_2012}, triblock copolymers with large polydispersity in middle B-blocks and small polydispersity in end A-blocks are examined. We assume that the middle B-blocks follows a Schulz-Zimm distribution with a $\text{PDI}_{\text{B}}$ $\sim$ 2.0 and the A-blocks follow a Poisson distribution with a $\text{PDI}_{\text{A}}$ $\sim$ 1.25~\cite{Mahanthappa_personal}. Figure~\ref{fig:FIG1}(b) shows the simulated phase diagram, where discrete symbols are the simulation results and the lines are guide to the eyes. The agreement between our simulation results and experimental observations is surprisingly good in the following aspects: (i) There is an obvious shift of the order-order transition boundaries towards larger volume fraction of the B-blocks ($f_{\text{B}}$) if compared with the monodisperse analogs. This result has also been confirmed by a recent SCFT calculation~\cite{Matsen_EPJ2013} and is quite similar to the influence of chain polydispersity on diblock copolymer microphase separation. For instance, the lamellar structures are only found when $f_{\text{B}}$ $>$ 0.5 in Figure~\ref{fig:FIG1}(b). It can be attributed to the polydispersity induced reduction in B-domain elasticity, which causes a tendency for the interface to curve towards polydisperse B-domain~\cite{ruzette_molecular_2006,beardsley_monte_2011,schmitt_polydispersity-driven_2012}. (ii) The composition window for irregular bicontinuous structures (BIC) is around 10$\%$ in our simulations, which is in very good agreement with experiment. Note that in order to make sure that the BIC structures we obtained are well equilibrated, we have preformed three repeated heating and cooling cycles on the structure. In each cycle, the system is firstly slowly heated up to temperature T = 3.0 and then cooled down back to T = 1.0. The heating/cooling rate is quite slow, with $\Delta$T = $\pm$0.5/500000 steps. This heating/cooling process is repeated for three times. After that, we find that the BIC structures remain. Interestingly, similar to SCFT results for diblock copolymers~\cite{matsen_polydispersity-induced_2007}, we also observe some two-phase coexistence regions on the order-order transition boundaries. For instance, coexistence of lamellar and bicontinuous structures (LAM+BIC) can be observed at $f_{\text{B}}$ = 0.5 as shown in Figure~\ref{fig:FIG1}(b). The agreements between our simulation results with both SCFT theory in long-chain limit (cf. Figure~\ref{fig:FIG1}(a)) and experiment (cf. Figure~\ref{fig:FIG1}(b)) indicate that our DPD simulations can represent the phase behaviors of realistic polymers.
\par
\begin{figure}
  \includegraphics[angle=0,width=10cm]{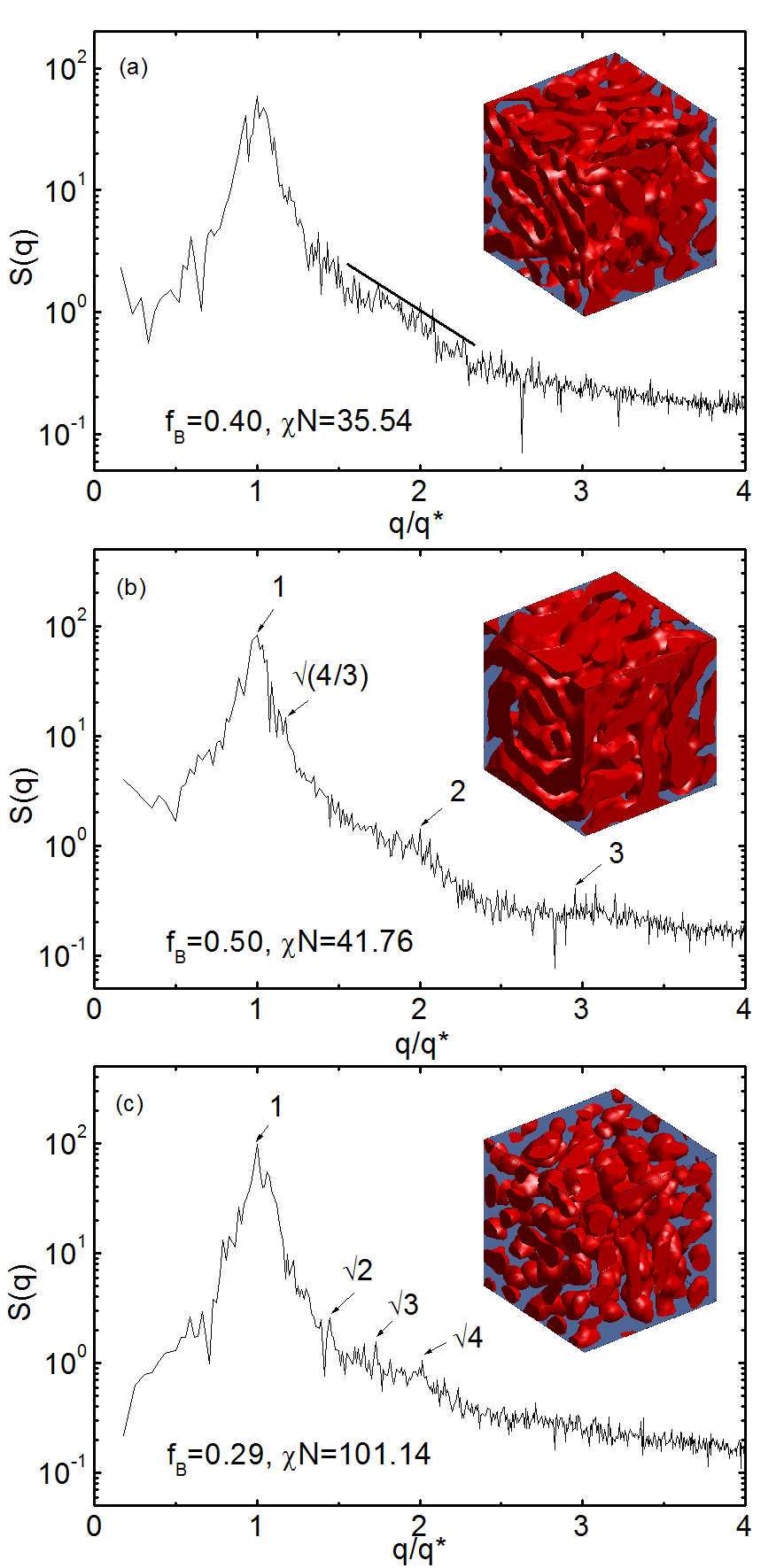}
  \caption{Structure factors and the corresponding mesostructures (insets): (a) irregular bicontinuous (BIC) structure formed in system 1-ABA-40 at $\chi$N = 35.54 with a broader peak in S($q$) indicated by the solid line; (b) coexistence of LAM+BIC phases formed in system 1-ABA-50 at $\chi$N = 41.76; (c) spherical phase formed in system 2-ABA-29 at $\chi$N = 101.14.}
  \label{fig:FIG3}
\end{figure}

One of the concerns in Refs.~\onlinecite{register_materials_2012,widin_unexpected_2012} is the influence of middle block polydispersity in the ABA triblock copolymers on the formation of the BIC structure. To clarify this, we studied a system with a medium polydispersity in the middle B-block (PDI$_{\text{B}}$ $\sim$ 1.5, obeying Schulz-Zimm distribution) but with monodisperse end A-blocks. The simulation results are presented in Figure~\ref{fig:FIG1}(c), which are in fact quite similar to those in Figure~\ref{fig:FIG1}(b): We can also observe ordered structures such as sphere (S), cylinder (C), irregular bicontinuous (BIC), and lamellar (LAM) structures. The composition window for the BIC structures is still around 10$\%$. Coexisting BIC and LAM structures are also found at $f_{\text{B}}$ = 0.5. The emergence of two-phase coexistence in the simulations of triblock copolymers is very similar to that in diblock copolymer system, which can be attributed to the fractionation of the polymers according to their molecular weights~\cite{matsen_polydispersity-induced_2007}. Typical structures of BIC, LAM+BIC, spherical structures and their corresponding structure factors are shown in Figure~\ref{fig:FIG3}. For the BIC structures (cf. Figure~\ref{fig:FIG3}(a)), there is a broad scattering peak (shoulder) indicated by a solid line after the primary peak, which is consistent with the experimental SAXS pattern~\cite{widin_unexpected_2012, widin_polydispersity-induced_2010}. For the coexisting LAM+BIC structures as shown in Figure~\ref{fig:FIG3}(b), we can observe characteristic peaks at $\vec{q}/\vec{q}^*$ = 1.0, 2.0, and 3.0 for lamellar structures, and characteristic double peaks at $\vec{q}/\vec{q}^*$ = 1.0 and $\sqrt{4/3}$ for gyroid structures and a broad peak at $\vec{q}/\vec{q}^*$ $\sim$ 2.0 for BIC structures. For the spherical structures shown in Figure~\ref{fig:FIG3}(c), peaks are found at $\vec{q}/\vec{q}^*$ = 1.0, $\sqrt2$, $\sqrt3$, and $\sqrt4$, which correspond to characteristic $bcc$ packing of spheres~\cite{ryan_ordered_2001}. These results suggest that suitably designed middle block polydispersity in ABA triblock copolymer can be used to engineer BIC structures.
\par
To examine the role played by the molecular weight distribution in the A-blocks on the formation of BIC structures, we focus on a system in which both A and B blocks obey Schulz-Zimm distribution with an equal PDI value of 1.5. Specifically, A and B blocks with different chain lengths are connected together in a random way to faithfully mimic the polydispersity distribution from a radical polymerization process. The phase diagram obtained in our simulations is presented in Figure~\ref{fig:FIG1}(d). Again, irregular bicontinuous structures with middle B-blocks as the minority component occupy about 10$\%$ of composition space. The position of this window is originally in the composition range of 0.35 $<$ $f_{\text{B}}$ $<$ 0.45 in Figure~\ref{fig:FIG1}(b) and (c), but now it is shifted to 0.3 $<$ $f_{\text{B}}$ $<$ 0.4 for both components being polydisperse, as shown in Figure~\ref{fig:FIG1}(d). A strikingly new result is that irregular bicontinuous structures are also observed at the right side of the phase diagram with high $f_{\text{B}}$  values 0.7 $<$ $f_{\text{B}}$ $<$ 0.8. Comparing to conventional monodisperse ABA triblock copolymer systems where bicontinuous gyroid structures can be found in only $\sim$ 3$\%$ of the composition space, the composition window for BIC structures in ABA triblock copolymer systems with synthetic polydispersity in both A and B blocks is enlarged to $\sim$ 20$\%$. Moreover, both the middle B and end A blocks have the same opportunity to form irregular network structures, which is particularly important for applications utilizing different functional properties arisen from distinct compositional blocks.
\par
%%%%%%%%%%%%
\subsection{Continuity of the bicontinuous structures}
%%%%%%%%%%%%
Bicontinuous structures at nanoscale is desirable in many applications, for example solar cells and batteries. The solid-state dye-sensitized solar cell~\cite{nature_cells_1991,bach_solid-state_1998} is viewed as one of the most promising nanotechnology-based hybrid photovoltaic system, where sensitizers are sandwiched between an $n$-type metal oxide and an organic hole transport material. The arrangement of these materials at the length scale of 10 nm will significantly affect the conversion of light to electrical energy. Recent experiments~\cite{crossland_bicontinuous_2009} showed that good connectivity of two separated phases formed by the self-assembly of block copolymers is conceptually ideal for the fabrication of hybrid solar cells. This is because that the excitons (electron-hole pairs), which are generated by photon absorption, could rapidly diffuse to the interface between bicontinuous separated domains allowing the charge separation and so as to generate a current. Therefore it is necessary to examine the continuity of both domains of the bicontinuous structure~\cite{register_materials_2012}.
\par
\begin{figure}
  \includegraphics[angle=0,width=12cm]{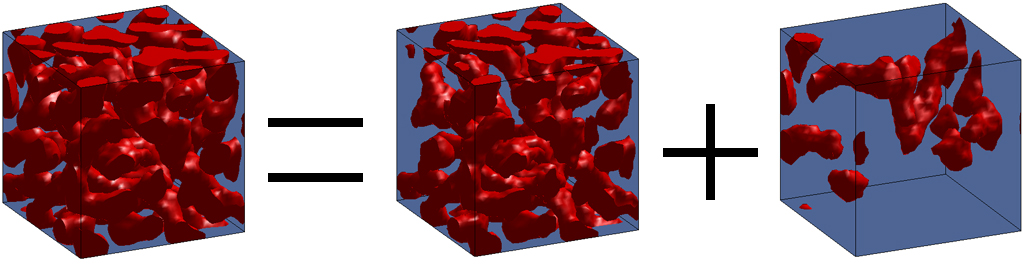}
  \caption{Decomposition of a BIC structure (left) formed in the system 1-ABA-375 at $\chi$N = 39.81 into one major domain (middle) and eight small domains (right). }
  \label{fig:FIG4}
\end{figure}

The continuity of the domain can be described by the domain integrity, which can be quantified by the number of separated domains in the system. Smaller number indicates better continuity. Therefore, we have calculated the number of separate domains ($N_{\text{domain}}$) for all obtained BIC structures in our simulations. Results are collected in Table~\ref{tb2:Ndomains} at various compositions $f_{\text{B}}$ for systems with different polydispersities. It demonstrates that the continuity of irregular bicontinuous structure is composition dependent. Optimum continuity can be found at the compositions close to the lamellar region in the phase diagram. For instance, $N_{\text{domain}}$ is found equal to 1 at $f_{\text{B}}$ = 0.45 for sample `1-ABA-45', at $f_{\text{B}}$ = 0.40 $\sim$ 0.45 for sample `2-ABA-40' and `2-ABA-0.45', and at $f_{\text{B}}$ = 0.375, 0.40, and 0.73 for `3-ABA' sample series, respectively. Comparison between series `1-ABA' (large polydispersity in middle B-block) and `2-ABA' (medium polydispersity in middle B-block) indicates that the increase in the polydispersity of middle B-block will decrease the domain integrity and result in more small domains (i.e.larger $N_{\text{domain}}$ value). Although the continuity of the bicontinuous structure is not perfect at other compositions (especially close to cylinder region in the phase diagram), the overall integrity of the domain is still quite good, only some small pieces of structures (small spheres) are found to be breaking apart from the main network structure. Domains in sample `1-ABA-375' at $f_{\text{B}}$ = 0.375 and \emph{$\chi$N} = 39.81 are shown in Figure~\ref{fig:FIG4}. The entire domain formed by B component is composed of 1 major continuous network structure and 8 small spherical or rod-like structures.
\par
%%%%%%%%%%%%%%%%%%%%%%%%%%%%%%%%%%%%%%%%%%%%%%%%%%%
\subsection{The domain size of the bicontinuous structures}
Since the domain size is another important factor for experimental applications, we have also calculated the domain size of the BIC structures for systems 3-ABA-375 (A$_{5}$B$_{6}$A$_{5}$, $f_{\text{B}}$ = 0.375), 3-ABA-40 (A$_{3}$B$_{4}$A$_{3}$, $f_{\text{B}}$ = 0.40) and 3-ABA-73 (A$_{2}$B$_{11}$A$_{2}$, $f_{\text{B}}$ = 0.73) at \emph{$\chi$N} $\approx$ 50, respectively. The domain size is calculated by $D = 2\pi/q^{*}$ following the experimental definition, where $q^{*}$ is the position of the primary peak in scattering function. Since the three systems have different chain lengths ($N_{total}$), the calculated domain sizes are normalized with $N_{total}$. The resulted values are $D/N_{total}$ = 0.498, 0.615 and 0.468 for three systems with $f_{\text{B}}$ = 0.375, 0.40 and 0.73, respectively. We observe that the BIC structures formed near to the composition window of lamellar structure possess a larger domain size than those near to the cylinder phase.
%%%%%%%%%%%%%%%%%%%%%%%%%%%%%%%%%%%%%%%%%%%%%%%%%%%

%%%%%%%%%%%%%%%%%%%%%%%%%%%%%%%%%%%%%%%%%%%%%%%%%%%%
\subsection{Origin of the stability of bicontinuous structures}
%%%%%%%%%%%%%%%%%%%%%%%%%%%%%%%%%%%%%%%%%%%%%%%%%%%%
As have been pointed out by Matsen and Bates~\cite{matsen_origins_1996,matsen_block_1997}, packing frustration plays an instrumental role in complex phase selection of block copolymers, especially in the cases of gyroid, perforated lamellar and double-diamond structures. In the self-assembly process of diblock copolymers, the molecules have a tendency to form domains of uniform thickness so that none of them are excessively stretched or compressed. In the cases of classical lamellar, cylinder and sphere phases, constant mean curvature model~\cite{CMC_1990} supplies good explanations for the stability of these ordered structures. Other than minimizing the interface area, the packing frustrations are also minimized since domains in these structures have a nearly uniform thickness~\cite{matsen_block_1997,bates_ordered_review_2009}. However, in the cases of gyroid, perforated lamellar, and double-diamond structures, there are unavoidable packing frustrations and curvature variations due to nonuniform thickness of the domains. Consequently the composition window for forming these complex structures is restricted to a narrow size. Disappearance of such complex structures at strong segregation can be also attributed to the exacerbated packing frustration~\cite{bates_ordered_review_2009}.
\par
\begin{figure}
  \includegraphics[angle=0,width=12cm]{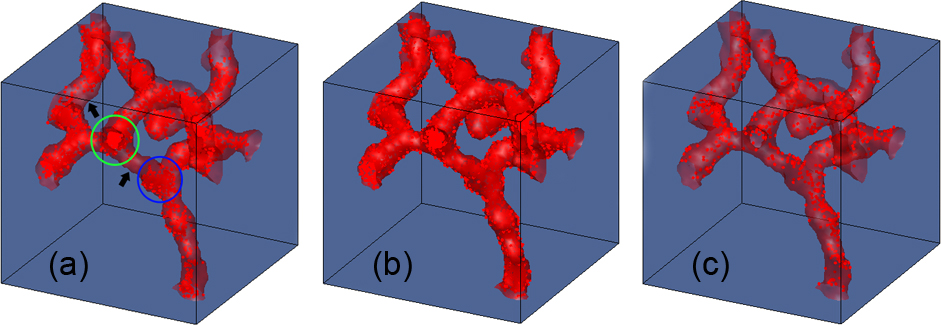}
  \caption{One of the isolated network structure formed by B component (isodensity surface in red) from the system 2-ABA-375 at $\chi$N = 62.95. The red dots in the structure show the distribution of (a) long B-blocks ($N_{\text{B}}$ $\geq$ 12), (b) intermediate B-blocks ($N_{\text{B}}$ = 4 $\sim$ 11), and (c) short B-blocks ($N_{\text{B}}$ $\leq$ 3). Blue and green circles in (a) indicate a threefold node structure and a fourfold node structure respectively. }
  \label{fig:FIG5}
\end{figure}

In the case of polydisperse systems, it is much easier for molecules to effectively fill the domain space due to the wide distribution of block chain length. Therefore the energy penalty due to packing frustration can be largely reduced. In order to directly illustrate the molecule distribution in BIC structures, we take `2-ABA-375' system as an example. Figure~\ref{fig:FIG5}(a) shows the distribution of the relatively long B-blocks ($N_{\text{B}}$ $\geq$ 12) in network structures formed by B component at $\chi$N = 62.95. In this irregular BIC structure, both threefold and fourfold network nodes are present, indicated by the blue and the green circles, respectively. Interestingly, long B-blocks are preferably located in the interior of these nodes due to the relatively larger space in the nodes. The accumulation of long blocks in the nodes of the BIC network structure is also expected to enhance the mechanical properties of the material. On the contrary, there are only a small portion of these long B-blocks found in the interior of the branching tubes of the network structures. The solid arrows in Figure~\ref{fig:FIG5}(a) indicate the positions where the tubes have smaller domain size and thus are free of long B-blocks. The domain space near these positions is filled by intermediate length B-blocks, which are actually almost homogeneously distributed over the BIC structure as indicated in Figure~\ref{fig:FIG5}(b). As expected, short length B-blocks are distributed at the interface, thus acting as surfactants (Figure~\ref{fig:FIG5}(c)). Therefore the packing frustration in the polydisperse system will be largely reduced via the selectively filling of long blocks in the network nodes. As a consequence, the stability of BIC structures formed by polydisperse ABA triblock copolymer can be enhanced in a wider composition window if compared with monodisperse systems. It is interesting to notice that similar stabilization effect by reducing the molecule packing frustration was also reported by adding homopolymer  ``additives'' inside the node interior~\cite{Escobedo_Macromolecules2007}.
%%%%%%%%%%%%%%%%%%%%%%%%%%%%%%%%%%%%%%%%%%%%%%%%%%%%
\section{CONCLUSIONS}
In summary, in order to explore the influence of polydispersity on the phase behavior of ABA triblock copolymers, the phase behavior of both monodisperse and polydisperse ABA triblock copolymers has been investigated using the particle-based dissipative particle dynamics simulation technique. In particular, the necessity of the narrow molecular weight distribution of end A-blocks, and the continuity in BIC structures are examined.
\par
Phase diagrams of monodisperse ABA triblock copolymers, polydisperse ABA triblock copolymers with polydispersity only presented in middle B-blocks and with synthetic polydispersity in both A and B blocks are constructed. Results of monodisperse systems show very good agreement with previous SCFT predictions: Ordered phases of cylinder, gyroid, and lamellar structures are found in the corresponding composition windows obtained by SCFT calculations. For a system mimicking the experimental system in Ref.~\onlinecite{widin_unexpected_2012} with large polydispersity in middle B-blocks (PDI$_{\text{B}}$ $\sim$ 2.0, Schulz-Zimm) and small polydispersity in end A-blocks (PDI$_{\text{A}}$ $\sim$ 1.25, Poisson), irregular bicontinuous structures are found over a composition window of 10$\%$ located between lamellar and cylinder phases. Ordered phases are shifted to larger compositions due to the polydispersity induced reduction in elasticity of B-domain, which causes a tendency for the interface to curve towards the B-domain. For the system with a smaller polydispersity in middle B-blocks (PDI$_{\text{B}}$ is reduced from 2.0 to 1.5), almost no changes are found in the phase diagram. The BIC structures are found to still occupy $\sim$ 10$\%$ of the composition space. The results of the system with an equal polydispersity in both A and B-blocks (PDI$_{\text{A}}$ = PDI$_{\text{B}}$ $\sim$ 1.5, Schulz-Zimm) show that the narrow distribution in the molecular weight of end A-blocks is not necessary for obtaining the BIC structure. The size of composition window for BIC network structures formed by B component is still $\sim$ 10$\%$ at the left side of the phase diagram, but more importantly, such BIC network structures formed by A component are also found at right side of the phase diagram, occupying also $\sim$ 10$\%$ of the composition space. This observation is particularly important for applications where bicontinuous domain structures formed by block copolymers are desirable. Such BIC structures are found to possess good continuity over the composition range where they form. Our study demonstrates that a simple radical polymerization process could be enough to synthesize polydisperse ABA triblock copolymers for fabricating bicontinuous materials. In addition, detailed analyses of the molecule distributions confirm that the stabilization of irregular bicontinuous structures over a wider composition range can be attributed to the reduction in molecule packing frustration with long B-blocks selectively distributed in the network nodes.
\section*{Acknowledgment}
This work is subsidized by the National Basic Research Program of China (973 Program, 2012CB821500), and supported by National Science Foundation of China (21204029, 21025416, 50930001) and by the Natural Science and Engineering Research Council (NSERC) of Canada.
%



%% References


%\bibliographystyle{model3-num-names}
%\bibliography{PDI-DIBLOCK-COPOLYMER}

%% not want to use model3-num-names.bst.





% \end{thebibliography}

\newpage
\begin{table}
  \caption{Molecular and morphological characteristics of simulated monodisperse ABA triblock copolymers. Symbol `$\&$' denotes that both phases are observed. $N_{\text{total}}$, $N_{\text{A}}$, and $N_{\text{B}}$ are the average length of the whole molecule, the length of A and B blocks, respectively.}
  \label{tb1:sample1}
  \begin{tabular}{C{2cm} C{1.5cm} C{1cm} C{1cm} C{1cm} C{1cm} C{1cm} C{1cm} C{2.5cm}}\hline\hline  Sample&$N_{\text{total}}$&$N_{\text{A}}$&$N_{\text{B}}$&PDI$_{\text{A}}$&PDI$_{\text{B}}$&PDI&$f_{\text{B}}$&Phase\\
  \hline
  mABA-25&8&3&2&1.00&1.00&1.00&0.25&HEX\\
  mABA-33&12&4&4&1.00&1.00&1.00&0.33&PL $\&$ G\\
  mABA-40&10&3&4&1.00&1.00&1.00&0.40&PL $\&$ LAM\\
  mABA-45&11&3&5&1.00&1.00&1.00&0.45&LAM\\
  mABA-50&12&3&6&1.00&1.00&1.00&0.50&LAM\\
  mABA-56&9&2&5&1.00&1.00&1.00&0.56&LAM\\
  mABA-60&10&2&6&1.00&1.00&1.00&0.60&PL $\&$ LAM\\
  mABA-67&12&2&8&1.00&1.00&1.00&0.67&PL\\
  mABA-75&8&1&6&1.00&1.00&1.00&0.75&HEX\\
  \hline
  \end{tabular}
\end{table}
\newpage

\newpage
\begin{table}
  \caption{Molecular and morphological characteristics of ABA triblock copolymers with large polydispersity in middle B-blocks (PDI$_{\text{B}}$ $\sim$ 2.0, Schulz-Zimm) and narrow distribution in end A-blocks (PDI$_{\text{A}}$ $\sim$ 1.25, Poisson). Symbol `$\&$' denotes that both phases are observed. LAM+BIC denotes the coexistence phase of LAM and BIC. $N_{\text{total}}^{\text{aver}}$, $N_{\text{A}}^{\text{aver}}$, and $N_{\text{B}}^{\text{aver}}$ are the average length of the whole molecule, the average length of A and B blocks, respectively.}
  \label{tb2:sample2}
  \begin{tabular}{L{2.2cm} C{1.5cm} C{1cm} C{1cm} C{1cm} C{1cm} C{1cm} C{1cm} C{2.5cm}}\hline\hline  Sample&$N_{\text{total}}^{\text{aver}}$&$N_{\text{A}}^{\text{aver}}$&$N_{\text{B}}^{\text{aver}}$&PDI$_{\text{A}}$&PDI$_{\text{B}}$&PDI&$f_{\text{B}}$&Phase\\
  \hline
  1-ABA-25&8.51&2.99&2.53&1.22&1.60&1.11&0.25&S\\
  1-ABA-29&14.03&4.98&4.06&1.16&1.47&1.08&0.29&C\\
  1-ABA-33&12.45&3.98&4.48&1.19&1.76&1.14&0.33&C\\
  1-ABA-375&16.41&4.98&6.44&1.16&1.82&1.16&0.375&BIC\\
  1-ABA-40&10.46&2.99&4.48&1.22&1.76&1.18&0.40&BIC\\
  1-ABA-45&11.44&2.99&5.47&1.22&1.80&1.21&0.45&BIC\\
  1-ABA-50&12.43&2.99&6.45&1.22&1.82&1.25&0.50&LAM+BIC\\
  1-ABA-56&9.45&1.99&5.47&1.25&1.80&1.29&0.56&LAM\\
  1-ABA-60&10.44&1.99&6.46&1.25&1.83&1.34&0.60&LAM\\
  1-ABA-67&12.41&1.99&8.42&1.25&1.86&1.41&0.67&PL $\&$ LAM\\
  \hline
  \end{tabular}
\end{table}
\newpage

\begin{table}
  \caption{Molecular and morphological characteristics of ABA triblock copolymers with polydisperse B-blocks (PDI$_{\text{B}}$ $\sim$ 1.5, Schulz-Zimm) and monodisperse end A-blocks. Symbol `$\&$' denotes that both phases are observed. LAM+BIC denotes the coexistence phase of LAM and BIC.}
  \label{tb3:sample3}
  \begin{tabular}{L{2.2cm} C{1.5cm} C{1cm} C{1cm} C{1cm} C{1cm} C{1cm} C{1cm} C{2.5cm}}\hline\hline  Sample&$N_{\text{total}}^{\text{aver}}$&$N_{\text{A}}^{\text{aver}}$&$N_{\text{B}}^{\text{aver}}$&PDI$_{\text{A}}$&PDI$_{\text{B}}$&PDI&$f_{\text{B}}$&Phase\\
  \hline
  2-ABA-17&12.35&5&2.35&1.00&1.56&1.02&0.17&S\\
  2-ABA-25&8.32&3&2.33&1.00&1.56&1.04&0.25&S\\
  2-ABA-29&14.12&5&4.12&1.00&1.49&1.04&0.29&S\\
  2-ABA-33&12.11&4&4.11&1.00&1.50&1.06&0.33&C\\
  2-ABA-375&16.04&5&6.04&1.00&1.49&1.07&0.375&C $\&$ BIC\\
  2-ABA-40&10.11&3&4.11&1.00&1.49&1.08&0.40&HEX $\&$ BIC\\
  2-ABA-45&10.98&3&4.98&1.00&1.49&1.10&0.45&BIC\\
  2-ABA-50&12.04&3&6.04&1.00&1.49&1.12&0.50&LAM+BIC\\
  2-ABA-56&8.99&2&4.99&1.00&1.49&1.15&0.56&LAM\\
  2-ABA-60&10.04&2&6.04&1.00&1.49&1.18&0.60&LAM\\
  2-ABA-67&12.00&2&8.00&1.00&1.49&1.22&0.67&LAM\\
  2-ABA-75&8.03&1&6.03&1.00&1.49&1.27&0.75&LAM+BIC\\
  \hline
  \end{tabular}
\end{table}
\newpage

\begin{table}
  \caption{Molecular and morphological characteristics of ABA triblock copolymers with an equal polydispersity in both A and B blocks, PDI$_{\text{A}}$ = PDI$_{\text{B}}$ $\sim$ 1.5 (Schulz-Zimm). LAM+BIC denotes the coexistence phase of LAM and BIC.}
  \label{tb4:sample4}
  \begin{tabular}{L{2.2cm} C{1.5cm} C{1cm} C{1cm} C{1cm} C{1cm} C{1cm} C{1cm} C{2.5cm}}\hline\hline  Sample&$N_{\text{total}}^{\text{aver}}$&$N_{\text{A}}^{\text{aver}}$&$N_{\text{B}}^{\text{aver}}$&PDI$_{\text{A}}$&PDI$_{\text{B}}$&PDI&$f_{\text{B}}$&Phase\\
  \hline
  3-ABA-25&8.35&3.10&2.15&1.45&1.39&1.15&0.25&C\\
  3-ABA-29&13.98&4.96&4.05&1.48&1.46&1.16&0.29&C\\
  3-ABA-33&12.19&4.06&4.06&1.47&1.47&1.16&0.33&BIC\\
  3-ABA-375&15.95&4.96&6.03&1.48&1.48&1.16&0.375&BIC\\
  3-ABA-40&10.25&3.10&4.05&1.45&1.46&1.15&0.40&BIC\\
  3-ABA-45&11.15&3.09&4.96&1.45&1.48&1.16&0.45&LAM+BIC\\
  3-ABA-50&12.20&3.10&6.01&1.45&1.48&1.17&0.50&LAM\\
  3-ABA-56&9.27&2.15&4.96&1.39&1.48&1.18&0.56&LAM\\
  3-ABA-60&10.34&2.15&6.03&1.39&1.48&1.20&0.60&LAM\\
  3-ABA-67&12.31&2.15&8.00&1.39&1.49&1.23&0.67&LAM+BIC\\
  3-ABA-73&15.28&2.15&10.97&1.39&1.49&1.27&0.73&BIC\\
  3-ABA-78&18.27&2.15&13.96&1.39&1.49&1.30&0.78&BIC\\
  \hline
  \end{tabular}
\end{table}
\newpage

\begin{table}
  \caption{Number of domains ($N_{\text{domain}}$) in BIC structures in simulated polydisperse systems.}
  \label{tb2:Ndomains}
  \begin{threeparttable}
  \begin{tabular}{L{3cm} L{2cm} C{1cm}} \hline\hline
%  \multicolumn{2}{c|}{PDI$_{A}$$ \sim $1.25 (PO),  PDI$_{B}$$ \sim $2.0 (SZ)} &
%  \multicolumn{2}{|c|}{PDI$_{A}$$=$1.00,  PDI$_{B}$$ \sim $1.50 (SZ)} &
%  \multicolumn{2}{|c}{PDI$_{A}$$ \sim $1.50 (SZ),  PDI$_{B}$$ \sim $1.50 (SZ)} \\
  System&$f_{\text{B}}$&$N_{\text{domain}}$\tnote{$a$}\\
  \cline{1-3}
  1-ABA-375&0.375&(3,9)\\
  1-ABA-40&0.40&(1,2)\\
  1-ABA-45&0.45&1\\
  \hline
  2-ABA-375&0.375&(1,4)\\
  2-ABA-40&0.40&1\\
  2-ABA-45&0.45&1\\
  \hline
  3-ABA-33&0.33&(1,2)\\
  3-ABA-375&0.375&1\\
  3-ABA-40&0.40&1\\
  3-ABA-73&0.73&1\\
  3-ABA-78&0.78&(2,6)\\
  \hline

  \end{tabular}

  \begin{tablenotes}
    \item[$a$] The numbers in parenthesis indicate the distribution of $N_{\text{domain}}$, where the first number is the smallest and the second number the largest values for $N_{\text{domain}}$ found in our simulations with different segregation strengths.
  \end{tablenotes}

\end{threeparttable}
\end{table}
\newpage

%%%%%%%%%%%%%%%% Include Figures
%%%%%%%%%%%%%%%% Include Figures
\newpage
\newpage
\newpage

\newpage

\newpage

\end{document}